\documentclass[aps,prm,twocolumn,superscriptaddress,floatfix,10pt]{revtex4}
\usepackage{graphicx}
\usepackage{amsmath}
\usepackage{color}
\usepackage{multirow}
\begin{document}

\title{Why do elastin-like polypeptides possibly have different solvation behaviors in water-ethanol and -urea mixtures?}

\author{Yani Zhao}
\affiliation{Max-Planck Institut f\"ur Polymerforschung, Ackermannweg 10, 55128 Mainz, Germany}
\author{Manjesh K. Singh}
\affiliation{Max-Planck Institut f\"ur Polymerforschung, Ackermannweg 10, 55128 Mainz, Germany}
\affiliation{Department of Mechanical Engineering, Indian Institute of Technology Kanpur, Kanpur 208016, India}
\author{Kurt Kremer}
\affiliation{Max-Planck Institut f\"ur Polymerforschung, Ackermannweg 10, 55128 Mainz, Germany}
\author{Robinson Cortes-Huerto}
\email{corteshu@mpip-mainz.mpg.de}
\affiliation{Max-Planck Institut f\"ur Polymerforschung, Ackermannweg 10, 55128 Mainz, Germany}
\author{Debashish Mukherji}
\email{debashish.mukherji@ubc.ca}
\affiliation{Stewart Blusson Quantum Matter Institute, University of British Columbia, Vancouver V6T 1Z4, Canada}

\date{\today}

\begin{abstract}
The solvent quality determines the collapsed or the expanded state of a polymer. For example, 
a polymer dissolved in a poor solvent collapses, whereas in a good solvent it opens up. While this standard understanding is generally valid, there are examples when a polymer collapses even in a mixture of two good solvents. This phenomenon, commonly known as co-non-solvency, is usually associated with a wide range of synthetic (smart) polymers. Moreover, recent experiments have shown that some biopolymers, such as elastin-like polypeptides (ELPs) that exhibit lower critical solution behavior $T_{\ell}$ in pure water, show co-non-solvency behavior in aqueous-ethanol mixtures.  
In this study, we investigate the phase behavior of elastin-like polypeptides (ELPs) in aqueous binary mixtures using molecular dynamics simulations of all-atom and complementary explicit solvent generic models. The model is parameterized by mapping the solvation free energy obtained from the all-atom simulations onto the generic interaction parameters.  For this purpose, we derive segment based (monomer level) generic parameters for four different peptides, namely proline (P), valine (V), glycine (G) and alanine (A), the first three constitute the basic building blocks of ELPs. Here we compare the conformational behavior of two ELP sequences, namely $-({\rm VPGGG})-$ and $-({\rm VPGVG})-$, in aqueous-ethanol and -urea mixtures. Consistent with recent experiments, we find that ELPs show co-non-solvency in aqueous-ethanol mixtures. Ethanol molecules have preferential binding with all ELP residues, with an interaction contrast of $6-8k_{\rm B}T$, and thus driving the coil-to-globule transition.
On the contrary, ELP conformations show weak variation in aqueous-urea mixtures. Our simulations suggest that the glycine residues dictate the overall behavior of ELPs in aqueous-urea, where urea molecules have a rather weak preferential binding with glycine, i.e., less than $k_{\rm B}T$. This weak interaction dilutes the overall effect of other neighboring residues and thus ELPs exhibit different conformational behavior in aqueous-urea in comparison to aqueous-ethanol mixtures. While the validation of the latter findings will require more detailed experimental investigation, the results presented here may provide a new twist to the present understanding of cosolvent interactions with peptides and proteins.
\end{abstract}

\maketitle

\section{Introduction}
\label{sec:intro}

Solvation of macromolecules in water and especially in a mixture of solvents is of central relevance for many areas of
chemical physics, polymer physics, soft matter science and materials 
research~\cite{cohen10natmat,mukherji14natcom,sissi14natcom,winnik15review,hoogenboom,mukherji17natcom,mukherji20arcmp}.
Indeed, solvation effects are the driving force underlying various macromolecular processes ranging from the 
responsiveness of hydrogels to external stimuli or concentration gradients of the
solvents (``smart polymers") to denaturation of proteins.
Furthermore, the relevant energy scale in these systems is of the order of thermal energy $k_{\rm B}T$,
with $k_{\rm B}$ is the Boltzmann constant and $T = 300$ K. Thus, the properties of macromolecules are
dictated by large conformational and compositional fluctuations.
Therefore, entropy (or generic physical laws) becomes as crucial as energy (or specific chemical details) for the study of these complex systems.
Admittedly, understanding this entropy-energy balance is at the heart of soft matter science~\cite{DeGennes,Grosbergbook,doibook}.

The flexibility of macromolecules provides a suitable platform for the tunable design of advanced
functional materials~\cite{ArmesLangmuir05,LutzJacs06,WinnikMacromole2018,MoehwaldAngewChem,Mukherji2019,Papadakis}. 
Furthermore, because of the carbon-based microscopic architectures they often create severe environmental problems. 
To circumvent this problem, recent interests have been directed toward the ``so-called" green chemistry~\cite{greenC}.
More specifically, making use of macromolecular structures that are bio-compatible~\cite{smartpolymer} and/or bio-degradable~\cite{Koberstein}, 
at the same time are also thermal~\cite{smartpolymer,Papadakis}, (co-)solvent 
\cite{Larson-Fredrickson,Tirrell,Winnik33,zhang2001,Winnik5,Sagle-etal-JACS131-9304-2009,Scherzinger,Walter,pnipam1,DzubiellaJCTC,Kyriakos},
and photo \cite{photoresp,Winnik7} responsive. While most of these systems are homopolymers, recent interests have been 
directed to a variety of copolymer architectures~\cite{Armes,cho,Tsitsilianis,weil,roy,mdelp,Roberts,Oliveira,Silva,bonduelle}.
Here, polypeptides and synthetic peptide-based polymers have attracted great interests~\cite{bonduelle,Roberts}.
In this context, elastin-like polypeptides (ELPs) represent a new class of stimuli-responsive synthetic polypeptides that show
vibrant phase behavior~\cite{Roberts,cho,weil,Glassman,Olsen}. 
Additionally, because of the biocompatible nature, ELPs
are used in many medicinal applications, such as the tissue scaffolding~\cite{Glassman}, cancer therapy~\cite{Saxena}, and protein purification~\cite{Hassouneh}.

ELPs, similar to many known smart polymers~\cite{cohen10natmat,mukherji14natcom,sissi14natcom,winnik15review,hoogenboom,mukherji17natcom,mukherji20arcmp}, 
exhibit rich and tunable phase diagrams in water~\cite{elp,weil,mdelp} and in aqueous
mixtures~\cite{Olsen}. Furthermore, because of the hydrogen bonding nature of the microscopic interaction, these
polymers are often water-soluble and, therefore, confer an expanded configuration of a chain for $T < T_{\ell}$, 
with $T_{\ell}$ being the lower critical solution temperature (LCST). 
When $T > T_{\ell}$ a certain number of bound water molecules are released from the polymer solvation volume
destabilizing an expanded polymer conformation~\cite{Grosbergbook,doibook}. 

\begin{figure*}[ptb]
\centering
\includegraphics[width=0.94\textwidth]{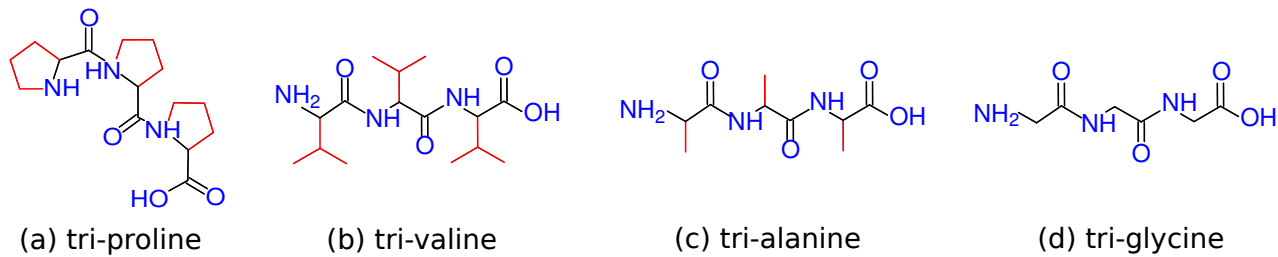}
        \caption{Schematic representation of the chemical structures of all four peptides,
        namely the (a) tri-proline, (b) tri-valine, (c) tri-alanine, and (d) tri-glycine. The hydrophilic
        parts are highlighted in blue, while the side chains are indicated in red.
} \label{schempep}
\end{figure*}

There has been considerable interest in studying polypeptides, ELPs and smart polymers 
using experiments~\cite{Tirrell,Winnik33,zhang2001,Winnik5,Sagle-etal-JACS131-9304-2009,Scherzinger,elp,Hellweg,Hietala,Laschewsky,Vishnevetskaya},
theory~\cite{Winnik5,Dudowicz,mukherji14natcom} and computer simulations~\cite{Walter,pnipam1,DzubiellaJCTC,mdelp,Silva}. 
While most studies of ELPs focus on  their behavior in water~\cite{cho,Hassouneh,elp,mdelp}, 
a recent study has also investigated the phase behavior of ELPs in aqueous ethanol mixtures~\cite{Olsen}.
The latter study has shown that starting from an expanded conformation of ELPs (for $T < T_{\ell}$) in water, addition of
ethanol molecules first collapses the chain. When ethanol concentration increases above a critical value, the chain again opens up. This coil-to-globule-to-coil transition in (miscible) binary mixtures is known as co-non-solvency, a name originally coined for
the study of polystyrene in cyclohexane-DMF mixtures~\cite{wolf} and later popularized for poly(N-isopropyl acrylamide) (PNIPAM) in aqueous alcohol mixtures~\cite{Tirrell,Winnik33}. The molecular origin of this
phenomenon has attracted intense debate in the literature. Here, various mechanisms have been proposed to be the
driving force for the phenomenon of co-non-solvency: namely the cooperativity effect~\cite{Winnik5,Winnik6,Kojima}, 
solvent-cosolvent interactions~\cite{Dudowicz,jia}, preferential cosolvent-monomer 
interaction~\cite{Larson-Fredrickson,Winnik4,DzubiellaJCTC,mukherji14natcom,Backes-etal-ACSML6-1042-2017}, and the
kosmotropic effect~\cite{Trappe}.

Even when the coil-to-globule transitions of polymers in aqueous-alcohol mixtures are prevalent cases, they also show interesting conformational behavior in water-urea mixtures. For example, PNIPAM also collapses under the influence of urea~\cite{Sagle-etal-JACS131-9304-2009}.
The origin of the urea induced collapse of PNIPAM was shown to be due to the urea induced bridging of two NIPAM monomers topologically far along a
polymer backbone~\cite{Sagle-etal-JACS131-9304-2009,Klitzing}.

In this work, we study and compare conformational behaviors of ELPs in water-ethanol and -urea mixtures using
molecular dynamics simulations of all-atom and complementary explicit solvent generic models.
Going beyond the simulation works dealing with implicit solvent generic model of ELPs 
under aqueous environments~\cite{Jayaraman}, our generic solvent models are derived 
by mapping solvation free energies obtained from the all-atom simulations onto
the generic explicit solvent model parameters for ELPs in binary solvents. We derive segment-based model parameters for four different amino acids
relevant for ELPs, namely Glycine (G), Alanine (A), Proline (P) and Valine (V) in
aqueous urea and aqueous ethanol mixtures.
The chemical structures for the trimers of these amino acids are shown in Fig. \ref{schempep}.
The model parameters are tested to reasonably reproduce the phase behavior of two ELP
sequences consisting of $({\rm VPGGG})$ and $({\rm VPGVG})$.
Note that here we do not attempt to address the secondary structures of polypeptides and/or 
copolymer of peptides \cite{pappu}. 
Moreover, because ELPs can be classified as intrinsically disordered proteins~\cite{Roberts}
their conformations can be described within the standard framework of polymer science~\cite{DeGennes,doibook},
which is the motivation behind this study

The remainder of the paper is organized as follows: the details of the all-atom simulations and the generic model parameterization is
presented in section~\ref{sec:method}. The conformations of generic polypeptides and ELPs in binary solution are shown in
section~\ref{sec:results}. Finally, we draw our conclusions in section~\ref{sec:summary}.

\section{Method and model}
\label{sec:method}

The generic model parameters are derived from solvation free-energy data obtained from all-atom simulations.
We also emphasize here that the generic model parameters for the ELPs are obtained at the
segment level, i.e., the parameters for different amino acids are obtained separately and then these
are used to simulate different ELP sequences \cite{Koberstein,dzubiella15jcp,Silva}. It should be noted that this approximation is 
in general valid for neutral monomers, which applies to the amino acids V, P, G and A. 
For charged monomers, it is necessary to refine the calculation of the solvation structure and 
the relative solvent-cosolvent coordination. Moreover, V, P, G and A are all very similar, and only the size of 
side carbon groups dictates their relative hydrophobicity, see Fig.\ \ref{schempep}. This similarity eliminates 
cross-correlation between different monomer units thus validating our segment-based approach \cite{Koberstein,dzubiella15jcp,Silva}.
Therefore, we start by describing the all-atom model used in this study.

\subsection{All-atom simulations}

All-atom simulations are performed using the GROMACS molecular dynamics package \cite{gromacs}.
These simulations are performed in the isobaric ensemble (NPT), where N is the number of particles,
P is the isotropic pressure, and T is the temperature. T = 300 K is set using a
velocity rescaling thermostat \cite{vrescale} with a coupling constant of 0.1 ps. 
Pressure is kept at 1 bar using a Parrinello-Rahman barostat~\cite{Parrinello-Rahman} with a coupling constant of
 2 ps. Electrostatics are treated with the particle mesh ewald (PME) method~\cite{PME}.
The interaction cutoff for the non-bonded interactions is chosen as 1.0 nm and the equations of
motions are integrated using the leap-frog integrator with a time step of  $\delta t =2$ fs.
All bonds were constrained with the LINCS~\cite{lincs}.

We investigate four trimers, namely tri-glycine, tri-alanine, tri-proline, and tri-valine, in aqueous urea
and aqueous ethanol mixtures (see peptide structures in Fig. \ref{schempep}).
The specific peptides are chosen because they constitute the monomeric building-blocks of ELPs.
Furthermore, we have only chosen trimers because the center monomer of a trimer gives
a reasonable estimate of the solvation structure and relative solvent-cosolvent coordination,
while not having to deal with conformation changes upon change in relative (co-)solvent compositions \cite{Silva}.

For the trimers, we have used the GROMOS43a1 force field \cite{Gromos43a1}. For water we use the SPC/E \cite{SPC3} model
and the Kirkwood-Buff (KB) derived force field of urea \cite{KBUWFF}. Note that the urea force-field was parameterized
on a GROMOS based model. Therefore, for consistency we have also used GROMOS43a1 for trimers.
We consider five different urea mole fractions $x_u$: 0.0382, 0.0809, 0.1292,
0.1844, and 0.2495. The total number of water and urea molecules are taken exactly the
same as in Ref.~\cite{debashish2012} that ensures solvent equilibrium within the simulation domain,
i.e., system sizes are large enough to neglect finite size effects.

Ethanol force field parameters are taken from~\cite{OPLS}.
The ethanol mole fractions $x_e$ are varied from pure water ($x_e = 0.0$) to $x_e = 0.25$. 
For aqueous ethanol solutions, we have taken a total number of 616 ethanol and 15528 water molecules 
at ethanol molar concentration $x_e$=0.0382, 1232 ethanol and 14000 water at $x_e$=0.0809, 1848 ethanol 
and 12456 water at $x_e$=0.1292, 2464 ethanol and 10896 water at $x_e$=0.1844, 3080 ethanol and 9263 water at $x_e$=0.2495. 

\subsection{Generic simulations}

Beyond the generic polymer model, we will describe the parameterization of: the
bulk binary solution, polymer-solvent interactions, and the model peptides in binary solutions.
Note that$-$ while we will describe the polymer model and polymer-solvent (polymer-water) interactions
in this section, polymer-cosolvent (polymer-urea and polymer-ethanol) parameterization will be described
whenever they are discussed in this manuscript.

\subsubsection{Polymer model}

To describe a polymer, we have used the well-known bead-spring model \cite{Kremer-Grest}.
In this model, monomers of a generic chain consist of Lennard-Jones (LJ) spheres.
Bond connectivity between adjacent monomers is introduced by a finite extensible nonlinear elastic (FENE) potential.
A bead-spring chain is solvated in mixtures of model water (solvent) and model ethanol or urea (cosolvent) molecules, 
also modeled as LJ spheres. The data is described in the units of LJ diameter $\sigma$, LJ energy $\varepsilon$ and mass $m$
of a monomer. This gives a time unit of $\tau = \sigma\sqrt{m/\varepsilon}$ and pressure $\varepsilon/\sigma^3$.
Therefore, following the simple protocol in which we map one amino acid onto one 
generic monomer. Given that all four peptides investigated in this study have very similar sizes, this mapping scheme is reasonable. 
Furthermore, if we look at the monomer units (see Fig. \ref{schempep}), they have typical sizes between 0.5-0.6 nm. This gives a
length scale mapping of about $1\sigma \approx 0.5$ nm.

The generic simulations are performed by using the ESPResSo++~\cite{espresso} and LAMMPS~\cite{lammps}
molecular dynamics packages.
Equations-of-motion are integrated with a velocity verlet algorithm with a time step $\delta t=0.005\tau$.
The damping coefficient $\Gamma$ of the Langevin thermostat is taken as $1.0\tau^{-1}$ to control the
temperature at $T = 1.0\epsilon/k_{\rm B}$.

Non-bonded monomers also interact with an attractive 6-12 LJ potential with $\sigma_m = 1\sigma$, $0.5\varepsilon \leq \varepsilon_m < 2.0 \varepsilon$,
and a cutoff $r_c = 2.5\sigma$. Details of monomer-monomer interactions will be described at the approprite section in this manuscript.
We have chosen a chain length $N_\ell = 50$ solvated in a
solvent box consisting of $5 \times 10^4$ particles.

\subsubsection{Bulk solution}

In this study we have used very simple spherically symmetric models for binary mixtures 
without any specific chemical details. Moreover, these model parameters give correct miscibilities 
known from the all-atom data of aqueous-ethanol \cite{Backes-etal-ACSML6-1042-2017} and aqueous-urea \cite{debashish2012}.\\ 

\noindent{\bf Water-ethanol mixtures:}
\begin{figure}[ptb]
\centering
\includegraphics[width=0.43\textwidth]{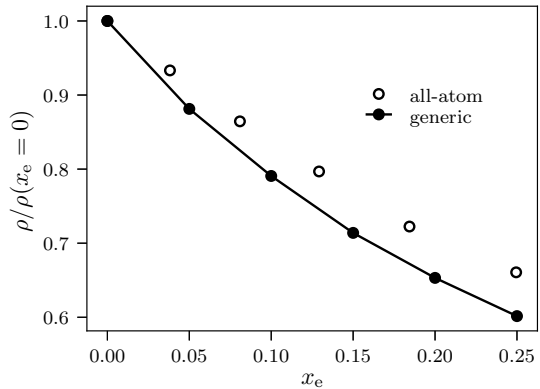}
        \caption{Normalized total number density $\rho/\rho(x_e=0)$ as a function of ethanol mole fraction $x_e$.
        Data is shown for all-atom and generic simulations. The line is drawn to guide the eye.
} \label{fig:densityPe}
\end{figure}
To model the bulk solution 
we consider that the size of water molecules ($\sim$0.28 nm) is typically half the size of the
peptides and the size of ethanol molecules ($\sim$0.50 nm) is about 1.8 times the size of water molecules.
Therefore, we choose the sizes of water and ethanol in our generic model to be $\sigma_{w} = 0.5~\sigma$ and 
$\sigma_{e} = 0.9 \sigma$, respectively. For $x_e = 0$ we choose number density $\rho_w = 5.5~\sigma^{-3}$,
while the SPC/E water has $\rho_w = 32$ nm$^{-3}$. The specific choice of $\rho_w$ is motivated by the fact that$-$ if we
choose $32$ nm$^{-3} = 5.5~\sigma^{-3}$, this will lead to $1\sigma \approx 0.55$ nm, which is consistent
with the length scale mapping described above. The generic water and ethanol molecules interact with each other
via the repulsive LJ interactions with $\varepsilon_{ij} = 1\varepsilon$, $\sigma_{ij} = \left(\sigma_{i}+\sigma_{j}\right)/2$
and a cutoff $2^{1/6}\sigma_{ij}$. With these parameters and for $x_e = 0$,
the typical pressure of the generic model is about $32\varepsilon/\sigma^3$. We have adjusted total $\rho$
with $x_e$ such that the pressure is kept constant. 
In Fig. \ref{fig:densityPe} we show a comparative plot of the normalized density $\rho/\rho(x_e = 0)$ as a function of $x_e$
between all-atom and generic simulations.
$\rho$ is consistent in both models as a function of $x_e$.\\

\begin{figure}[ptb]
\centering
\includegraphics[width=0.43\textwidth]{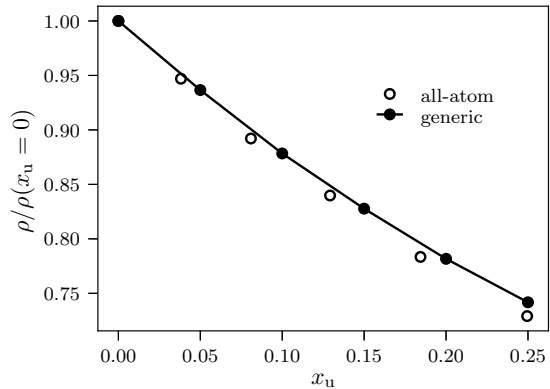}
	\caption{Same as Fig. \ref{fig:densityPe}, however, for aqueous urea mixtures.
} \label{densityP}
\end{figure}

\noindent{\bf Water-urea mixtures:}
Similar to the parameterization of water-ethanol mixtures, we have also parameterized aqueous urea mixtures.
For this purpose we consider the size of urea molecules ($\sim$0.42 nm) to be about 1.5 times the size of water molecules.
Therefore, we choose $\sigma_{w} = 0.5~\sigma$ and $\sigma_{u} = 0.75\sigma$.
The generic water and urea molecules interact with each other
via the repulsive LJ ineterations with $\varepsilon_{ij} = 1\varepsilon$, $\sigma_{ij} = \left(\sigma_{i}+\sigma_{j}\right)/2$
and a cutoff $2^{1/6}\sigma_{ij}$. We have adjusted total $\rho$ with $x_u$ such that the pressure is kept constant 
at $32\varepsilon/\sigma^3$. In Fig. \ref{densityP} we show normalized density $\rho/\rho(x_u = 0)$ as a function of $x_u$ 
comparing all-atom and generic simulations.

\section{Results and discussions}
\label{sec:results}

\subsection{Elastin-like polypeptides in aqueous ethanol}

Before describing the conformation of ELPs in aqueous ethanol mixtures, we will first start our discussion by 
describing the polymer-(co-)solvent interaction.

\subsubsection{Good solvent case of peptide-solvent interactions}

One of the most important factors in modelling ELPs in solution is to properly capture the relative 
affinities of different peptides in pure water (solvent). 
In this context, to investigate the coil-to-globule transition of ELPs in aqueous ethanol mixtures, we consider that the ELP chain
is under good solvent condition in pure water, i.e., when $T < T_{\ell}$. To obtain model parameters that reasonably satisfy 
the good solvent condition of ELPs in pure water, we have estimated the possible number of hydrogen bonds (${\rm H-bond}$) between 
an amino acid and water molecules $N_w$ from the all-atom simulations. H$-$bonds are calculated using the standard
GROMACS subroutine, where a H$-$bond exists when the donor-acceptor distance is $\leq 0.35$ nm
and the acceptor-donor-hydrogen angle is $\leq 30^{\circ}$.
The data is summarized in the column 3 of Table \ref{tab:hbond_e}.
\begin{table}[ptb]
        \caption{Number of hydrogen bonds between the center monomer of the trimers with water $N_w$ and with ethanol $N_e$.
	Data is shown for different ethanol mole fractions $x_e$.
        }
\centering
\begin{tabular}{|c |c |c|c |c |c |c |c |}
\hline
\multicolumn{2}{|c|}{$x_e$}&0.0000 &0.0382 &0.0809 &0.1292 & 0.1844 &0.2495  \\
\hline
\multirow{2}{*}{P}
& $N_w$ & 0.806& 0.519 & 0.346 & 0.200 & 0.134 & 0.107 \\
& $N_e$ & --   & 0.109 & 0.163 & 0.196 & 0.195 & 0.199 \\
\hline
\multirow{2}{*}{V}
& $N_w$ & 1.617& 0.987 & 0.601 & 0.449 & 0.311 & 0.247 \\
& $N_e$ & --   & 0.501 & 0.706 & 0.881 & 0.872 & 0.915 \\
\hline
\multirow{2}{*}{A}
& $N_w$ & 1.660& 1.097 & 0.756 & 0.524 & 0.369 & 0.257 \\
& $N_e$ & --   & 0.534 & 0.748 & 0.870 & 0.971 & 1.060 \\
\hline
\multirow{2}{*}{G}
& $N_w$ & 2.094  & 1.244 & 0.920 & 0.617 & 0.463 & 0.341 \\
& $N_e$ & --  & 0.419 & 0.655 & 0.865 & 0.946 & 1.029 \\
\hline
\end{tabular}
\label{tab:hbond_e}
\end{table}
It can be appreciated that$-$
\begin{equation*}
N_w^{\rm proline} < N_w^{\rm valine} \approx N_w^{\rm alanine} < N_w^{\rm glycine}. \nonumber \\
\end{equation*}
To model the above described relative affinities, we have used LJ interaction parameters described in Table \ref{tab:epssig_e}. 
\begin{table*}[ptb]
	\footnotesize
        \caption{A table listing the Lennard-Jones (LJ) length $\sigma_{ij}$ and energy $\epsilon_{ij}$ parameters for all 
	pairs of particles in the generic model of valine (V), proline (P), alanine (A) and glycine (G) in water-ethanol mixtures.}
	\setlength{\tabcolsep}{0.8em}
\begin{tabular}{|c |c |c|c |c |c |c |c |c |c|c |c |c|}
\hline
\multirow{2}{*}{} & \multicolumn{2}{c|}{water} & \multicolumn{2}{c|}{ethanol} & \multicolumn{2}{c|}{G} & \multicolumn{2}{c|}{P} & \multicolumn{2}{c|}{A}& \multicolumn{2}{c|}{V} \\
\cline{2-13}
&$\sigma_{ij}$ &$\epsilon_{ij}$ &$\sigma_{ij}$ &$\epsilon_{ij}$ & $\sigma_{ij}$ &$\epsilon_{ij}$ & $\sigma_{ij}$ &$\epsilon_{ij}$
        & $\sigma_{ij}$ &$\epsilon_{ij}$ & $\sigma_{ij}$ &$\epsilon_{ij}$ \\
\hline
water & 0.500 & 1.000 & 0.700 & 1.000 & 0.750 & 0.680 & 0.750 & 0.480 & 0.750 & 0.670& 0.750& 0.500\\
\hline
ethanol&0.700 & 1.000 & 0.900 & 1.000 & 0.950 & 2.150& 0.950 &1.900 & 0.950 & 2.110& 0.950 & 2.100\\
\hline
G & 0.750 & 0.680 & 0.950 & 2.150 & 1.000 & 0.500& 1.000 & 0.500& 1.000 & 0.500& 1.000 & 0.500\\
\hline
P & 0.750 & 0.480 & 0.950 & 1.900 & 1.000 & 0.500& 1.000 & 0.500& 1.000 & 0.500& 1.000 & 0.500\\
\hline
A & 0.750 & 0.670 & 0.950 & 2.110 & 1.000 & 0.500& 1.000 & 0.500& 1.000 & 0.500& 1.000 & 0.500\\
\hline
V & 0.750 & 0.500 & 0.950 & 2.100 & 1.000 & 0.500& 1.000 & 0.500& 1.000 & 0.500& 1.000 & 0.500\\
\hline
\end{tabular}
\label{tab:epssig_e}
\end{table*}
These specific parameter choices ensure that a chain consisting of model amino acids 
remains expanded in pure solvent with attractive affinity with the solvent (water) molecules via hydrogen bonding,
as is known from the ELP chain conformation below its $T_{\ell}$~\cite{Olsen}.
Furthermore, it should be noted that this parameter space is not restricted and similar solvation
conditions can also be achieved with different sets of parameters as long as relative monomer-monomer and monomer-solvent
interactions are considered consistently.

\subsubsection{Peptide-ethanol interactions}

For the parameterization of the model to study ELPs in aqueous ethanol mixtures, we map the solvation free energies $G_p$
obtained from the all-atom simulations of amino acids onto the generic model~\cite{mukherji14natcom,mukherji17natcom}. 
To obtain $G_p$ we have used the Kirkwood-Buff theory of solutions~\cite{kbi}, which connects the fluctuation in the
grand canonical ensemble $\mu$VT with the pair-wise solution structure of complex fluids using the ``so called" 
Kirkwood-Buff integral (KBI),
\begin{equation}
\begin{split}
        G^{\mu VT}_{ij}&= V\left[\frac{\langle N_iN_j\rangle-\langle N_i\rangle\langle N_j\rangle}
	{\langle N_i\rangle\langle N_j\rangle}-\frac{\delta_{ij}}{\langle N_i\rangle}\right]  \\
        &=4\pi\int_0^\infty [g_{ij}^ {\rm {\mu}VT}(r)-1]r^2 \text{d}r.
\end{split} \label{kkbbi}
\end{equation}
Here $G^{\mu VT}_{ij}$ and $g_{ij}^{\rm {\mu}VT}(r)$ are the KBI and the radial distribution function 
between $i$ and $j$ solution components in the $\mu$VT ensemble,
respectively. $\mu$ is the chemical potential. $\langle\cdot \rangle$ gives the ensemble average, 
$\delta_{ij}$ is the Kronecker delta, and $N_i$ is the number of particles of type
$i$. In the thermodynamic limit $G^{\mu {\rm VT}}_{ij} \approx G^{{\rm NVT}/{\rm NPT}}_{ij}$.
Here, however, we obtain $G_{ij}$ from $4\pi \int_0^{r_{\rm o}}\left[g_{ij}(r)-1\right]r^2 {\rm d}r$ with $r_{\rm o} = 1.5$ nm. 
Note that within the finite simulation domains this is a safe approximation given that the typical correlation length in the 
aqueous systems is within the range $1.5-2.0$ nm~\cite{pnipam1}. Furthermore, all system sizes are chosen 
to be the same as our earlier works that ensure well converged $G_{ij}$ \cite{debashish2012,pnipam1,Backes-etal-ACSML6-1042-2017}. 
There are also more accurate methods to obtain $G_{ij}$ directly from fluctuations~\cite{robin2016,Petris}, 
here we take the rather simple route of estimating $G_{ij}$ from the convergence of KBIs.

$G_{ij}$ can be used to derive the solvation free energy. When peptides ($p$) under infinite dilution are disolved in a
mixture of water ($w$) and ethanol ($e$), the shift in the solvation free energy ${\triangle G}_p$ can be
calculated using,
\begin{equation}
\underset{\rho_p\rightarrow 0}{\lim} \left(\frac{\partial\triangle G_p}{\partial x_e} \right )_{p, T}
        = \frac{RT(\rho_w+\rho_e)^2} {\eta}(G_{pw}-G_{pe}).
\end{equation}
Here $\rho_i$ is the number density and $\eta= \rho_w+\rho_e+\rho_w\rho_e \left(G_{ww}+G_{ee}-2G_{we}\right)$.
Additionally, the preferential solvation parameter $\left(G_{ww}+G_{ee}-2G_{we}\right)$ gives the direct measure of
the miscibility in bulk (binary) solution. Here we find $\left(G_{ww}+G_{ee}\right) \approx 2G_{we}$
for both all-atom and generic simulations over a full range of $x_e$, indicating almost perfect miscibility
as shown earlier \cite{pnipam1,Backes-etal-ACSML6-1042-2017}.

\begin{figure}[ptb]
\centering
\includegraphics[width=0.46\textwidth]{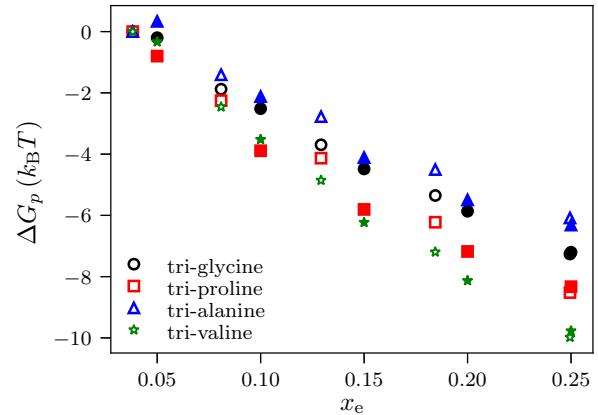}
        \caption{The shift in solvation free energy $\triangle G_p$ per monomer as a function of ethanol mole fraction $x_e$.
        Data is shown for four different trimers, namely tri-glycine, tri-proline, tri-alanine and tri-valine.
        The all-atom data is shown by empty symbols and the filled symbols corresponds to the generic model. $\triangle G_p$ is calculated with respect to the center monomer of a trimer.
	Note that the center monomer is chosen because it has correct solvation structure as known from a monomer in a long chain.
}\label{solvationwe}
\end{figure}

In Fig. \ref{solvationwe} we show $\triangle G_p$ with changing $x_e$ obtained from all-atom simulations (see empty symbols).
We tune monomer-cosolvent (peptide-ethanol) interactions in our generic model to reproduce this shift in $\triangle G_p$, as
seen from the open symbols in Fig. \ref{solvationwe}. The details of model parameters are described in Table \ref{tab:epssig_e}. 
Furthermore, as demonstrated in Table \ref{tab:hbond_e} we find$-$ $N_e^{\rm proline} < N_e^{\rm valine} < N_e^{\rm alanine} \approx N_e^{\rm glycine}$.
Therefore, we incorporate above conditions in our generic model via:
$\epsilon_{w}^{\rm proline} < \epsilon_{w}^{\rm valine} < \epsilon_{w}^{\rm alanine} < \epsilon_{w}^{\rm glycine}$
and $\epsilon_{e}^{\rm proline} < \epsilon_{e}^{\rm valine} < \epsilon_{e}^{\rm alanine} < \epsilon_{e}^{\rm glycine}$,
see Table \ref{tab:epssig_e}.
Fig. \ref{solvationwe} also shows that ethanol has a preferential interaction with all amino acids, which is 
about 6-8 $k_{\rm B}T$, more than the peptide-water interactions.

\subsubsection{Conformation of elastin-like polypeptides in aqueous ethanol}

\begin{figure}[ptb]
\centering
\includegraphics[width=0.45\textwidth]{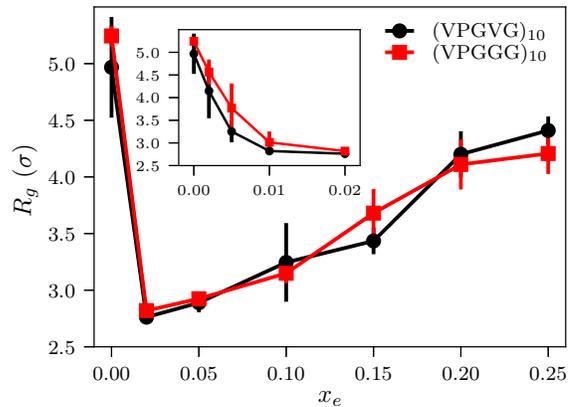}
        \caption{Gyration radius $R_g$ of two ELP sequences, namely $-$(VPGVG)$-$ and $-$(VPGGG)$-$
	with ten repeat units, as a function of ethanol mole fraction $x_e$. 
	Data is shown from the simulations using the parameters presented in Table \ref{tab:epssig_e}. Inset: Detail of the interval $0.00\le x_e\le 0.02$.
}\label{rgELPew}
\end{figure}

Using the generic model of peptides in aqueous ethanol described above, we will now investigate 
ELP conformations in aqueous-ethanol mixtures. In this context, ELPs are one of the most intriguing classes of 
polymers that are genetically engineered having the properties of polymer random coil 
and at the same time are bio-compatible because of their amino acid-based monomeric building blocks. 
Here, ELPs usually have the sequence VPG$-$X$-$G, where X can be any amino acid except proline~\cite{elp}.
Because of the dominant H$-$bond nature of the interaction between amino acid and water molecules, 
ELPs show LCST behavior. Here, $T_{\ell}$ can be tuned by varying ELP sequences. 
For example, $T_{\ell} \approx 300-305$ K for X = Valine~\cite{Urry}, while $T_{\ell} \approx 305-310$ K 
for X = Glycine (i.e., more hydrophilic residue)~\cite{Urry,Olsen}. This is identical to the typical LCST based copolymers, where $T_{\ell}$ can be
tuned by changing hydrophobic or hydrophilic units along the backbone~\cite{dzubiella15jcp,Koberstein,Silva,Oliveira}.
Therefore, for this study we investigate two sequences, namely $-({\rm VPGVG})-$ and $-({\rm VPGGG})-$, for $T < T_{\ell}$. 
Note that both these systems remain expanded at around $T \approx 300$ K \cite{Urry,Olsen}, where our generic models are parameterized.
Using the default parameters (see Table \ref{tab:epssig_e}) ELP conformations are studied 
with changing ethanol concentrations. In Fig. \ref{rgELPew} we show $R_{\rm g}$ for two sequences as a function of $x_e$. 
It can be seen that$-$ starting from an expanded chain in pure water ($x_e = 0$) increasing $x_e$ first collapses a 
chain between $0.05 < x_e < 0.15$, upon further increase of $x_e \geq 0.15$ an ELP chain re-opens.
This coil-to-globule-to-coil transition, often referred to as co-non-solvency~\cite{Tirrell,Winnik33},
is a well known phenomenon of standard smart polymers \cite{pnipmam,Ishihara,Scherzinger,Ohkura,Hiroki}.
Moreover, a recent experiment has also shown that ELPs can exhibit co-non-solvency in aqueous ethanol mixtures.
In this context, our results are in good agreement with the experimental data~\cite{Olsen}. 
Furthermore, not only is the conformational behavior observed in simulations consistent with experiments, 
but the window of collapse is also in reasonable agreement with the experimental measurement 
at $300$ K, i.e., $0.05 < x_e < 0.14$ ~\cite{Olsen}.

While the microscopic origin of the co-non-solvency phenomenon is a matter on intense debate, it has been previously shown that the 
preferential binding of the better solvents (in this case ethanol) with the monomers drives the polymer 
collapse~\cite{Larson-Fredrickson,Winnik4,DzubiellaJCTC,mukherji14natcom}. 
When a small amount of ethanol is added into the aqueous solution of ELPs, these molecules preferentially bind to 
more than one amino acid to reduce the binding free energy. This leads to a typical case where a certain number of 
ethanol molecules form sticky contacts between different amino acids thus initiating ELP collapse.
Furthermore, it was also discussed that this collapse can not be explained within the standard 
Flory-Huggins like mean-filed picture, where the solvent-monomer and cosolvent-monomer interactions are
dominant in comparison to the bulk solution $\chi$ parameter \cite{mukherji15jcp}. This also justifies our
choice of spherically symmetric particles representing bulk solution components, which only 
requires $\chi \simeq 0$ as known from the most common solvent mixtures where co-non-solvency is observed \cite{Tirrell}.
Additionally, the interaction interaction between pure solvent and pure cosolvent with monomer 
should be $\sim 4k_{\rm B}T$ to observe co-non-solvency~\cite{pnipam1}. 
When this contrast reduces to $ \le 2k_{\rm B}T$, no co-non-solvency is observed~\cite{pnipam2}.
In this context, we find that all four amino acids have very strong preferential binding (6$-$8 $k_{\rm B}T$) 
with ethanol in comparison to water (see Fig. \ref{solvationwe}). Therefore, it is expected that an
ELP shows the standard co-non-solvency in aqueous ethanol. We would like to mention that 
if the residue X is replaced with A, a $-$(VPGAG)$-$ sequence will also show similar conformational behavior as 
shown in Fig. \ref{rgELPew} because A has a very similar contrast of $\triangle G_p$ in aqueous ethanol as G or V (see Fig. \ref{solvationwe}).

\begin{figure}[ptb]
\centering
\includegraphics[width=0.46\textwidth]{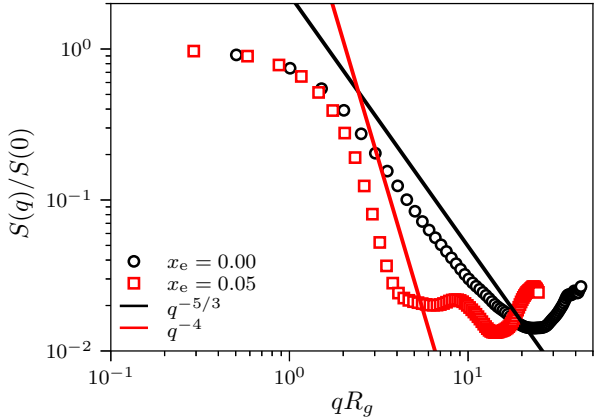}
	\caption{Normalized single chain form factor $S(q)/S(0)$ of a $-$(VPGVG)$-$ sequence 
	for $x_e = 0.0$ (black) and $x_e = 0.05$ (red). Lines are power law fits with 
	scaling exponents $q^{-5/3}$ and $q^{-4}$. 
} \label{sqvpgvg}
\end{figure}

To further investigate ELP collapse, we have also calculated the single chain form factor $S(q)$
in Fig. \ref{sqvpgvg}. While the data for the pure solvent (i.e., $x_e = 0.0$) shows $q^{-5/3}$ scaling
as expected for a good solvent chain, data for $x_e = 0.05$ shows $q^{-4}$ behavior until $qR_{\rm g} \sim 4.0$
and then deviates for $qR_{\rm g} > 4.0$.
It is noticeable that the ELP for $x_e = 0.05$ does not show a perfect sphere scattering as expected 
from $S(q)$ of a collapsed polymer globule~\cite{DeGennes,Grosbergbook,doibook}. 
In this context, it is worthwhile to mention that even when a polymer collapses 
in a mixture of two good solvents, it is not a standard poor solvent collapse dictated by 
depletion interactions~\cite{mukherji17natcom}. Instead, solvent quality becomes better-and-better with 
increasing $x_e$, as evident from the ever decreasing variation of $\triangle G_p$ with $x_e$ (see Fig. \ref{solvationwe}).
This implies that even though a polymer remains collapsed under the influence of binary good solvents,
it consists of good solvent blobs. Ideally speaking, for long chains, a cross-over from $q^{-4}$ to $q^{-5/3}$
gives the typical blob size~\cite{pnipam3}. Here, however, our chain length is rather small 
and, therefore, we do not observe any cross-over scaling.

Having discussed the conformation of ELPs in aqueous ethanol mixtures, we now want to  investigate a broader 
implication of ELP conformations in other binary mixtures, such as the aqueous urea mixtures. 
In this context, it should be mentioned that one of the most studied polymers that show co-non-solvency in aqueous alcohol mixtures is PNIPAM. 
Here, PNIPAM not only collapses in aqueous alcohol, but also shows an interesting coil-to-globule transition in aqueous 
urea solution~\cite{Sagle-etal-JACS131-9304-2009,Klitzing}
In these studies, it was shown that the strong H$-$bonding between the hydrophilic 
group of NIPAM monomers and urea drives the collapse of a chain. Here, the mechanism of polymer collapse was 
shown to be driven by urea molecules forming sticky contacts between distant monomers far along the polymer backbone.
Therefore, it is worth investigating if urea can also confer collapse of an ELP, which is the motivation behind the next section. 

\subsection{Elastin-like polypeptides in aqueous urea}

\begin{figure}[ptb]
\centering
\includegraphics[width=0.40\textwidth]{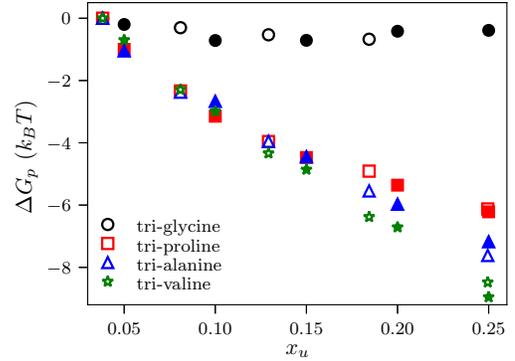}
        \caption{The shift in solvation free energy $\triangle G_p$ as a function of urea mole fraction $x_u$.
        Data is shown for four different trimers, namely tri-glycine, tri-proline, tri-alanine and and tri-valine.
        The all-atom data is shown by empty symbols and the filled symbols correspond to the generic model. The all-atom data for tri-glycine is taken from Ref.\ \cite{debashish2012}.
}\label{solvationG}
\end{figure}

Urea is a well known denaturant for proteins or peptides~\cite{ureadenature1}. 
Here, however, we will investigate the effect of urea on the possible folding transition of good solvent 
ELPs in pure water. To mimic the good solvent case, we have taken the same monomer-solvent (amino acid-water) 
parameters as presented in Table \ref{tab:epssig_e}.
Consistently, the generic monomer-cosolvent (amino acid-urea) interaction parameters are obtained by 
mapping ${\triangle G}_p$ onto the all-atom data, using the same protocol presented in Section 3.1.2.
\begin{table*}[ptb]
\footnotesize
	\caption{Same as Table \ref{tab:epssig_e}, however, for peptides in water and urea mixtures.}
\setlength{\tabcolsep}{0.8em}
\begin{tabular}{|c |c |c|c |c |c |c |c |c |c|c |c |c|}
\hline
\multirow{2}{*}{} & \multicolumn{2}{c|}{water} & \multicolumn{2}{c|}{urea} & \multicolumn{2}{c|}{G} 
	& \multicolumn{2}{c|}{P} & \multicolumn{2}{c|}{A}& \multicolumn{2}{c|}{V} \\
\cline{2-13}
&$\sigma_{ij}$ &$\epsilon_{ij}$ &$\sigma_{ij}$ &$\epsilon_{ij}$ & $\sigma_{ij}$ &$\epsilon_{ij}$ & $\sigma_{ij}$ &$\epsilon_{ij}$
        & $\sigma_{ij}$ &$\epsilon_{ij}$ & $\sigma_{ij}$ &$\epsilon_{ij}$ \\
\hline
water & 0.500 & 1.000 & 0.625 & 1.000 & 0.750 & 0.680 & 0.750 & 0.480 & 0.750 & 0.670 & 0.750 & 0.500\\
\hline
urea & 0.625 & 1.000 & 0.750 & 1.000 & 0.875 & 1.100& 0.875 &1.350 & 0.875 & 1.800 & 0.875 & 1.760\\
\hline
G & 0.750 & 0.680 & 0.875 & 1.100 & 1.000 & 0.500 & 1.000 & 0.500 & 1.000 & 0.500 & 1.000 & 0.500 \\
\hline
P & 0.750 & 0.480 & 0.875 & 1.350 & 1.000 & 0.500 & 1.000 & 0.500 & 1.000 & 0.500 & 1.000 & 0.500 \\
\hline
A & 0.750 & 0.670 & 0.875 & 1.800 & 1.000 & 0.500 & 1.000 & 0.500 & 1.000 & 0.500 & 1.000 & 0.500 \\
\hline
V & 0.750 & 0.500 & 0.875 & 1.760 & 1.000 & 0.500 & 1.000 & 0.500 & 1.000 & 0.500 & 1.000 & 0.500 \\
\hline
\end{tabular}
\label{epssiguw}
\end{table*}
In Table \ref{epssiguw} we present the full list of generic monomer (co-)solvent interactions. This parameter set also ensures that
the shift in $\triangle G_p$ is well reproduced in the generic model as known from the all-atom simulations (see Fig. \ref{solvationG}). 
Fig. \ref{solvationG} also shows that the relative preferentiability of urea with glycine is almost
negligible as indicated by $|G_p(x_u = 0) - G_p(x_u = 0.2495)| < k_{\rm B}T$.
The all atom data for glycine is taken from our earlier work \cite{debashish2012},
which is consistent with other experimental \cite{mvalue} and simulation data \cite{rnetz}.
For the other three amino acids the shift is between 6-8 $k_{\rm B}T$, thus showing that the
urea interaction is highly preferred with these amino acids in comparison to water.
It should also be noted that urea molecules not only interact with an amino acid with preferential 
H$-$bond, but also interact with the hydrophobic residues of different 
amino acids (see red parts in Figs. \ref{schempep}(a-c)) via van der Waals interactions~\cite{Berne}.

\begin{figure}[ptb]
\centering
\includegraphics[width=0.45\textwidth]{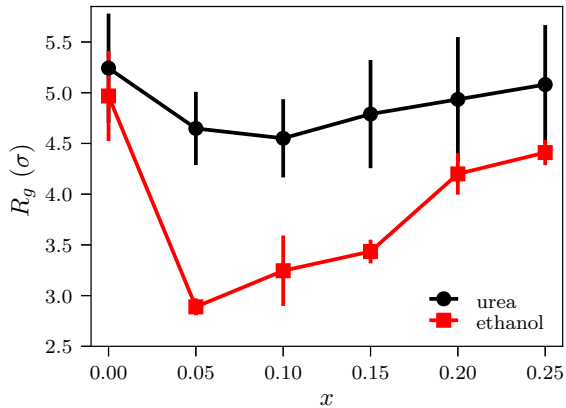}
	\caption{Gyration radius $R_g$ of $-$(VPGVG)$-$ sequence with ten repeat units 
	in aqueous urea (black) and aqueous ethanol (red) mixtures 
	as a function of their molar concentrations $x$.}\label{elpgood}
\end{figure}

In Fig. \ref{elpgood} we show the conformational behavior of a ELP sequence of $-$(VPGVG)$-$ in aqueous-urea mixtures. 
For comparison, we have also included the data for aqueous-ethanol mixtures, see Fig. \ref{rgELPew}.
It can be appreciated by the black data set in Fig. \ref{elpgood} that $-$(VPGVG)$-$ shows weak swelling-collapse-swelling 
behavior in aqueous-urea in comparison to the aqueous-ethanol mixtures (see red data set in Fig. \ref{elpgood}).
It is still important to mention that valine and proline residues have 6-8 $k_{\rm B}T$ interaction contrast
and thus a chain should collapse~\cite{pnipam3}. Here, however, the effect is diluted because of the dominant effect of the 
glycine residues that are in the majority and almost have no preferentiability between water and urea (see Fig. \ref{solvationG}).
In this context, it is important to mention that it requires a certain concentration of hydrophobic 
residues along the backbone to initiate the polymer collapse, which is more than 50\% in most cases \cite{dzubiella15jcp,Silva}. Therefore,
while $-$(VPGVG)$-$ still shows weak collapse, conformation of $-$(VPGGG)$-$ shows no noticeable change with $x_u$.
While validating this scenario would require detailed experimental work, this already highlights that if glycine can have
strong preference with a cosolvent one can observe a standard coil-to-globule transition as shown earlier in Section 3.2. 

We would also like to clarify why glycine has higher interaction strength with ethanol than with urea.
Here, we observe that ethanol molecules bind with a glycine via a preferential H$-$bond. This leads to a typical case where 
the hydrophobic $-$CH$_2$CH$_3$ part of ethanol gets exposed to the bulk water forming large hydrophobic patches along the ELP backbone. 
These hydrophobic patches can then confer a collapsed conformation through the hydrophobic interactions. However, in the case of urea and
water there are only dominant H$-$bonds, leading to complete mixing of all solution species.

\section{Conclusions}
\label{sec:summary}

We have derived explicit solvent generic models to study conformation of peptides and ELPs in aqueous mixtures. 
The parameterization procedure is done by mapping the solvation free energies obtained from the all-atom simulations onto the generic model interaction parameters
with changing cosolvent concentration. The mapping is performed at the monomer level for different peptides, namely proline (P), valine (V), glycine (G) and alanine (A),
where the first three are typical building blocks of ELPs. These models are used to study the conformational behavior of 
ELPs in aqueous ethanol and aqueous urea mixtures. Note that by conformation, we only mean the size of an ELP without 
attempting to describe the protein secondary structures. We find that ELPs show fascinating co-non-solvency behavior in aqueous ethanol, 
as observed in recent experimental work~\cite{Olsen}. We rationalize this result in terms of the preferential ethanol 
interactions with all peptide residues of ELPs. By contrast, the degree of collapse of ELPs in aqueous urea is rather weak.
This distinct behavior can be attributed to the difference in glycine interactions in aqueous ethanol in comparison to aqueous urea. 
Some of our results are in direct agreement with the existing experimental data. Furthermore, we also make predictions. 
Indeed, we present the first set of simulations on ELPs in binary mixtures giving direct evidence towards a more robust and 
tunable phase behavior of bio-compatible systems. Therefore, these results may provide a new directions to the 
advanced materials design.

\section*{Acknowledgments}
Y.Z. thanks Tiago E. de Oliveira for the useful discussions regarding the all-atom force field. This work has partially been supported
by European Research Council under the European Union's Seventh Framework
Programme (FP7/2007-2013)/ERC Grant Agreement No. 340906-MOLPROCOMP. Y.Z. and R. C.-H. gratefully acknowledge partially funding
by the Deutsche Forschungsgemeinschaft (DFG, German Research Foundation) - Project number 233630050 - TRR 146. D.M. thanks Canada First Research Excellence Fund for financial support.
We thank Carlos M. Marques, David Ng and Nancy C. Forero-Martinez for stimulating discussions. We further acknowledge
Carlos M. Marques, Torsten Stuehn, Nancy C. Forero-Martinez, and Jasper Michels for their very useful comments on the manuscript.


\begin{thebibliography}{25}

\bibitem{cohen10natmat}
Cohen-Stuart, M. A.; Huck, W. T. S.; Genzer, J.; M\"uller, M.; Ober, C.; Stamm, M.;  Sukhorukov, G. B.; Szleifer, I.; Tsukruk, V. V.; Urban, M.; Winnik, F.; Zauscher, S.;  Luzinov, I.; Minko, S. Emerging applications of stimuli-responsive polymer materials. 
{\it Nat. Materials} {\bf 2010}, {\it 9}, 101.

\bibitem{mukherji14natcom}
Mukherji, D.; Marques, C. M.; Kremer, K. Polymer collapse in miscible good solvents is a generic phenomenon driven by preferential adsorption. 
{\it Nat. Commun.} {\bf 2014}, {\it 5}, 4882.

\bibitem{sissi14natcom}
de Beer, S.; Kutnyanszky, E.; Sch\"on, P. M.; Vancso, G. J.; M\"user, M. H. 
Solvent-induced immiscibility of polymer brushes eliminates dissipation channels. 
{\it Nat. Commun.} {\bf 2014}, {\it 5}, 3781.

\bibitem{winnik15review}
Halperin, A.; Kr\"oger, M.; Winnik, F. M.
Poly(N-isopropylacrylamide) phase diagrams: fifty years of research.
{\it Angew. Chem. Int. Ed.} {\bf 2015}, {\it 54(51)}, 15342-15367.

\bibitem{hoogenboom}
Zhang, Q. R.; Hoogenboom, R.
Polymers with upper critical solution temperature behavior in alcohol/water solvent mixtures.
{\it Prog. Pol. Science} {\bf 2015}, {\it 48}, 122-142.

\bibitem{mukherji17natcom}
Mukherji, D.; Marques, C. M.; St\"uhn, T.; Kremer, K.
Depleted depletion drives polymer swelling in poor solvent mixtures.
{\it Nat. Commun.} {\bf 2017}, {\it 8}, 1374.


	\bibitem{mukherji20arcmp}
		Mukherji, D.; Marques, C. M.; Kremer, K.
		 Smart responsive polymers: Fundamentals and design principles. {\it Ann. Rev. Cond. Mat. Phys.} {\bf 2020}, {\it 11}, 271.


\bibitem{DeGennes}
De Gennes, P. G. Soft matter. {\it Rev. Mod. Phys.} {\bf 1992}, {\it 64(3)}, 645.

\bibitem{Grosbergbook}
Grosberg, A. Y.; Khokhlov, A.R. Statistical Physics of Macromolecules, AIP Press: 1994; New York.

\bibitem{doibook}
Doi, M. Soft matter physics. Oxford University Press: 2013; New York.

\bibitem{MoehwaldAngewChem}
Donath, E.; Sukhorukov, G. B.; Caruso, F.; Davis, S. A.; M\"ohwald, H.  
Novel Hollow Polymer Shells by Colloid-Templated Assembly of Polyelectrolytes. {\it Angew. Chem.} {\bf 1998}, {\it 110(16)}, 2323-2327.

\bibitem{ArmesLangmuir05}
Li, C.; Buurma, N.J.; Haq, I.; Turner, C.; Armes, S.P.; Castelletto, V.; Hamley, I.W.; Lewis, A. L. Synthesis and characterization of biocompatible, thermoresponsive ABC and ABA triblock copolymer gelators. {\it Langmuir} {\bf 2005}, {\it 21(24)}, 11026-11033.

\bibitem{LutzJacs06}
Lutz, J. F.; Akdemir, \"O.; Hoth, A. Point by point comparison of two thermosensitive polymers exhibiting a similar LCST: is the age of poly (NIPAM) over? {\it J. Am. Chem. Soc.} {\bf 2006}, {\it 128(40)}, 13046-13047.

\bibitem{WinnikMacromole2018}
Tao, D.; Feng, C.; Lu, Y.; Cui, Y.; Yang, X.; Manners, I.; Winnik, M.A.; Huang, X. 
Self-seeding of block copolymers with a $\pi$-conjugated oligo (p-phenylenevinylene) segment: a versatile route toward monodisperse fiber-like nanostructures. {\it Macromolecules} {\bf 2018}, {\it 51(5)}, 2065-2075.

\bibitem{Mukherji2019}
Mukherji, D.; Watson, M. D.; Morsbach, S.; Schmutz, M.; Wagner, M.; Marques, C. M.; Kremer, K. Soft and Smart: Co-nonsolvency-Based Design of Multiresponsive Copolymers. {\it Macromolecules} {\bf 2019}, {\it 52}, 3471.

\bibitem{Papadakis}
Papadakis, C. M.; M\"uller-Buschbaum, P.; Laschewsky, A. Switch It Inside-Out: ``Schizophreni" Behavior of All Thermoresponsive UCST-LCST Diblock Copolymers. {\it Langmuir} {\bf 2019}. DOI: 10.1021/acs.langmuir.9b01444.

\bibitem{greenC}
Anastas, P. T.; Kirchhoff, M. M. Origins, current status, and future challenges of green chemistry. {\it Acc. Chem. Res.} {\bf 2002}, {\it 35(9)}, 686-694.

\bibitem{smartpolymer}
Weber, C.; Hoogenboom, R.; Schubert, U. S. Temperature responsive bio-compatible polymers based on poly(ethylene oxide) and poly(2-oxazoline)s. 
{\it Prog. Polym. Sci.} {\bf 2012}, {\it 37(5)}, 686-714.

\bibitem{Koberstein}
Samanta, S.; Bogdanowicz, D. R.; Lu, H. H.; Koberstein, J. T. Polyacetals: Water-Soluble, pH-Degradable Polymers with Extraordinary Temperature Response. {\it Macromolecules} {\bf 2016}, {\it 49(5)}, 1858-1864.

\bibitem{Larson-Fredrickson}
Magda, J. J.; Fredrickson, G. H.; Larson, R. G.; Helfand, E. Dimensions of a polymer chain in a mixed solvent. {\it Macromolecules} {\bf 1988}, {\it 21(3)}, 726-732.

\bibitem{Tirrell}
Schild, H. G.; Muthukumar, M.; Tirrell, D. A. Cononsolvency in mixed aqueous solutions of poly (N-isopropylacrylamide). {\it Macromolecules} {\bf 1991}, {\it 24(4)}, 948-952.

\bibitem{Winnik33}
Winnik, F.M.; Ringsdorf, H.; Venzmer, J. Methanol-water as a co-nonsolvent system for poly (N-isopropylacrylamide). {\it Macromolecules} {\bf 1990}, {\it 23(8)}, 2415-2416.

\bibitem{zhang2001}
Zhang, G.; Wu, C. Reentrant coil-to-globule-to-coil transition of a single linear homopolymer chain in a water/methanol mixture. {\it Phys. Rev. Lett.} {\bf 2001}, {\it 86(5)}, 822.

\bibitem{Winnik5}
Tanaka, F.; Koga, T.; Winnik, F. M. Temperature-responsive polymers in mixed solvents: competitive hydrogen bonds cause cononsolvency. {\it Phys. Rev. Lett.} {\bf 2008}, {\it 101(2)}, 028302.

\bibitem{Sagle-etal-JACS131-9304-2009}
Sagle, L. B.; Zhang, Y.; Litosh, V. A.; Chen, X.; Cho, Y.; Cremer, P. S.  Investigating the hydrogen-bonding model of urea denaturation. {\it J. Am. Chem. Soc.} {\bf 2009}, {\it 131(26)}, 9304-9310.

\bibitem{Scherzinger}
Scherzinger, C.; Lindner, P.; Keerl, M.; Richtering, W. Cononsolvency of Poly(N, N-diethylacrylamide)(PDEAAM) and Poly(N-isopropylacrylamide)(PNIPAM) Based Microgels in Water/Methanol Mixtures: Copolymer vs Core-Shell Microgel. {\it Macromolecules} {\bf 2010}, {\it 43(16)}, 6829-6833.

\bibitem{Walter}
Walter, J.; Sehrt, J.; Vrabec, J.; Hasse, H. Molecular dynamics and experimental study of conformation change of poly (N-isopropylacrylamide) hydrogels in mixtures of water and methanol. {\it J. Phys. Chem. B} {\bf 2012}, {\it 116(17)}, 5251-5259.

\bibitem{pnipam1}
Mukherji, D.; Kremer, K. Coil-globule-coil transition of pnipam in aqueous methanol: Coupling all-atom simulations to semi-grand canonical coarse-grained reservoir. {\it Macromolecules} {\bf 2013}, {\it 46(22)}, 9158-9163.

\bibitem{DzubiellaJCTC}
Heyda, J.; Muzdalo, A.; Dzubiella, J. Rationalizing polymer swelling and collapse under attractive cosolvent conditions. {\it Macromolecules} {\bf 2013}, {\it 46(3)}, 1231-1238.

\bibitem{Kyriakos}
Kyriakos, K.; Philipp, M.; Lin, C. H.; Dyakonova, M.; Vishnevetskaya, N.; Grillo, I.; Zaccone, A.; Miasnikova, A.; Laschewsky, A.; M\"uller-Buschbaum, P.; Papadakis, C. M.  Quantifying the interactions in the aggregation of thermoresponsive polymers: the effect of cononsolvency. {\it Macromol. Rapid Commun.} {\bf 2016}, {\it 37(5)}, 420-425.

\bibitem{photoresp}
Dai, S.; Ravi, P.; Tam, K. C. Thermo-and photo-responsive polymeric systems. {\it Soft Matter} {\bf 2009}, {\it 5(13)}, 2513-2533.

\bibitem{Winnik7}
Ishii, N.; Mamiya, J. I.; Ikeda, T.; Winnik, F. M. Solvent induced amplification of the photoresponsive properties of $\alpha$, $\omega$-di-[4-cyanophenyl-4'-(6-hexyloxy)-azobenzene]-poly(N-isopropylacrylamide) in aqueous media. {\it Chem. Comm.} {\bf 2011}, {\it 47(4)}, 1267-1269.

\bibitem{Armes}
Li, C.; Madsen, J.; Armes, S. P.; Lewis, A. L. A new class of biochemically degradable, stimulus-responsive triblock copolymer gelators. {\it Angew. Chem.} {\bf 2006}, {\it 45(21)}, 3510-3513.

\bibitem{cho}
Cho, Y.; Zhang, Y.; Christensen, T.; Sagle, L. B.; Chilkoti, A.; Cremer, P. S.  Effects of Hofmeister anions on the phase transition temperature of elastin-like polypeptides. {\it J. Phys. Chem. B} {\bf 2008}, {\it 112(44)}, 13765-13771.

\bibitem{Tsitsilianis}
Tsitsilianis, C. Responsive reversible hydrogels from associative ``smart" macromolecules. {\it Soft Matter} {\bf 2010}, {\it 6(11)}, 2372-2388.

\bibitem{weil}
Wu, Y.; Pramanik, G.; Eisele, K.; Weil, T. Convenient approach to polypeptide copolymers derived from native proteins. {\it Biomacromolecules} {\bf 2012}, {\it 13(6)}, 1890-1898.

\bibitem{roy}
Roy, D.; Brooks, W. L.; Sumerlin, B. S. New directions in thermoresponsive polymers. {\it Chem. Soc. Rev.} {\bf 2013}, {\it 42(17)}, 7214-7243.

\bibitem{mdelp}
Li, N. K.; Quiroz, F. G.; Hall, C. K.; Chilkoti, A.; Yingling, Y. G. Molecular description of the LCST behavior of an elastin-like polypeptide. {\it Biomacromolecules} {\bf 2014}, {\it 15(10)}, 3522-3530.

\bibitem{Roberts}
Roberts, S.; Dzuricky, M.; Chilkoti, A. Elastin-like polypeptides as models of intrinsically disordered proteins. {\it FEBS lett.} {\bf 2015}, {\it 589(19PartA)}, 2477-2486.

\bibitem{Oliveira}
de Oliveira, T. E.; Mukherji, D.; Kremer, K.; Netz, P. A. Effects of stereochemistry and copolymerization on the LCST of PNIPAm. {\it J. Chem. Phys.} {\bf 2017}, {\it 146(3)}, 034904.

\bibitem{Silva}
De Silva, C. C.; Leophairatana, P.; Ohkuma, T.; Koberstein, J. T.; Kremer, K.; Mukherji, D. Sequence transferable coarse-grained model of amphiphilic copolymers. {\it J. Chem. Phys.} {\bf 2017}, {\it 147(6)}, 064904.

\bibitem{bonduelle}
Bonduelle, C. Secondary structures of synthetic polypeptide polymers. {\it Polym. Chem.} {\bf 2018}, {\it 9(13)}, 1517-1529.

\bibitem{Glassman}
Glassman, M. J.; Avery, R. K.; Khademhosseini, A.; Olsen, B. D. Toughening of thermoresponsive arrested networks of elastin-like polypeptides to engineer cytocompatible tissue scaffolds. {\it Biomacromolecules} {\bf 2016}, {\it 17(2)}, 415-426.

\bibitem{Olsen}
Mills, C. E.; Ding, E.; Olsen, B. D. Cononsolvency of elastin-like polypeptides in water/alcohol solutions. {\it Biomacromolecules} {\bf 2019}, {\it 20(6)}, 2167-2173.

\bibitem{Saxena}
Saxena, R.; Nanjan, M. J. Elastin-like polypeptides and their applications in anticancer drug delivery systems: a review. {\it Drug delivery} {\bf 2015}, {\it 22(2)}, 156-167.
\bibitem{Hassouneh}
Hassouneh, W.; Christensen, T.; Chilkoti, A. Elastin-like polypeptides as a purification tag for recombinant proteins. {\it Curr. Protoc. Protein Sci.} {\bf 2010}, {\it 61(1)}, 6-11.
\bibitem{elp}
Meyer, D. E.; Chilkoti, A. Genetically encoded synthesis of protein-based polymers with precisely specified molecular weight and sequence by recursive directional ligation: examples from the elastin-like polypeptide system. {\it Biomacromolecules} {\bf 2002}, {\it 3(2)}, 357-367.

\bibitem{Hellweg}
Karg, M.; Hellweg, T. New ``smart" poly (NIPAM) microgels and nanoparticle microgel hybrids: Properties and advances in characterisation. {\it Curr. Opin. Colloid Interface Sci.} {\bf 2009}, {\it 14(6)}, 438-450.

\bibitem{Hietala}
Hietala, S.; Nuopponen, M.; Kalliom\"aki, K.; Tenhu, H. Thermoassociating poly (N-isopropylacrylamide) A-B-A stereoblock copolymers. {\it Macromolecules} {\bf 2008}, {\it 41(7)}, 2627-2631.
\bibitem{Laschewsky}
Arot\c{c}ar\'ena, M.; Heise, B.; Ishaya, S.; Laschewsky, A. Switching the inside and the outside of aggregates of water-soluble block copolymers with double thermoresponsivity. {\it J. Am. Chem. Soc.} {\bf 2002}, {\it 124(14)}, 3787-3793.
\bibitem{Vishnevetskaya}
Vishnevetskaya, N. S.; Hildebrand, V.; Niebuur, B. J.; Grillo, I.; Filippov, S. K.; Laschewsky, A.; M\"uller-Buschbaum, P.; Papadakis, C. M. ``Schizophrenic" Micelles from Doubly Thermoresponsive Polysulfobetaine-b-poly (N-isopropylmethacrylamide) Diblock Copolymers. {\it Macromolecules} {\bf 2017}, {\it 50(10)}, 3985-3999.

\bibitem{Dudowicz}
Dudowicz, J.; Freed, K. F.; Douglas, J. F. Communication: Cosolvency and cononsolvency explained in terms of a Flory-Huggins type theory. {\it J. Chem. Phys.} {\bf 2015}, {\it 143}, 131101.

\bibitem{wolf}
Wolf, B. A.; Willms, M. M. Measured and calculated solubility of polymers in mixed solvents: Co-nonsolvency. {\it Die Makromolekulare Chemie: Macromol. Chem. Phys.} {\bf 1978}, {\it 179(9)}, 2265-2277.

\bibitem{Kojima}
Kojima, H.; Tanaka, F.; Scherzinger, C.; Richtering, W. Temperature dependent phase behavior of PNIPAM microgels in mixed water/methanol solvents. {\it J. Pol. Sci. B} {\bf 2013}, {\it 51(14)}, 1100-1111.

\bibitem{Winnik6}
Tanaka, F.; Koga, T.; Kojima, H.; Xue, N.; Winnik, F. M. Preferential adsorption and co-nonsolvency of thermoresponsive polymers in mixed solvents of water/methanol. {\it Macromolecules} {\bf 2011}, {\it 44(8)}, 2978-2989.

\bibitem{jia}
Jia, D.; Zuo, T.; Rogers, S.; Cheng, H.; Hammouda, B.; Han, C. C. Re-entrance of Poly (N, N-diethylacrylamide) in D2O/d-Ethanol Mixture at 27$^\circ$ C. {\it Macromolecules} {\bf 2016}, {\it 49(14)}, 5152-5159.

\bibitem{Backes-etal-ACSML6-1042-2017}
Backes, S.; Krause, P.; Tabaka, W.; Witt, M. U.; Mukherji, D.; Kremer, K.; von Klitzing, R. Poly (N-isopropylacrylamide) microgels under alcoholic intoxication: When a lcst polymer shows swelling with increasing temperature. {\it ACS Macro Lett.} {\bf 2017}, {\it 6(10)}, 1042-1046.

\bibitem{Winnik4}
Winnik, F. M.; Ottaviani, M. F.; Bossmann, S. H.; Garcia-Garibay, M.; Turro, N. J.  Consolvency of poly (N-isopropylacrylamide) in mixed water-methanol solutions: a look at spin-labeled polymers. {\it Macromolecules} {\bf 1992}, {\it 25(22)}, 6007-6017.

\bibitem{Trappe}
Bischofberger, I.; Calzolari, D. C.; Trappe, V. Co-nonsolvency of PNiPAM at the transition between solvation mechanisms. {\it Soft Matter} {\bf 2014}, {\it 10(41)}, 8288-8295.

\bibitem{Klitzing}
Micciulla, S.; Michalowsky, J.; Schroer, M. A.; Holm, C.; von Klitzing, R.; Smiatek, J. Concentration dependent effects of urea binding to poly (N-isopropylacrylamide) brushes: a combined experimental and numerical study. {\it Phys. Chem. Chem. Phys.} {\bf 2016}, {\it 18(7)}, 5324-5335.

\bibitem{Jayaraman}
Prhashanna, A.; Taylor, P. A.; Qin, J.; Kiick, K. L.; Jayaraman, A. Effect of Peptide Sequence on the LCST-Like Transition of Elastin-Like Peptides and Elastin-Like Peptide-Collagen-Like Peptide Conjugates: Simulations and Experiments. {\it Biomacromolecules} {\bf 2019}, {\it 20(3)}, 1178-1189.

\bibitem{pappu}
Holehouse, A. S.; Pappu R. V.
Collapse Transitions of Proteins and the Interplay Among Backbone, Sidechain, and Solvent Interactions.
{\it Ann. Rev. Biophys.} {\bf 2018}, {\it 47}, 19-39.

\bibitem{dzubiella15jcp}
Schulz, B.; Chudoba, R.; Heyda, J.; Dzubiella, J. Tuning the critical solution temperature of polymers by copolymerization. {\it J. Chem. Phys.} {\bf 2015}, {\it 143(24)}, 243119.

\bibitem{gromacs}
Pronk, S.; P\'all, S.; Schulz, R.; Larsson, P.; Bjelkmar, P.; Apostolov, R.; Shirts, M. R.; Smith, J. C.; Kasson, P. M.; van der Spoel, D.; Hess, B.; Lindahl, E.  GROMACS 4.5: a high-throughput and highly parallel open source molecular simulation toolkit. {\it Bioinformatics} {\bf 2013}, {\it 29(7)}, 845-854.
\bibitem{vrescale}
	Bussi, G.; Donadio, D.; Parrinello, M. {\it J. Chem. Phys.} {\bf 2007} {\it 126}, 014101.
Berendsen, H. J.; Postma, J. V.; van Gunsteren, W. F.; DiNola, A. R. H. J.; Haak, J. R. Molecular dynamics with coupling to an external bath. {\it J. Chem. Phys.} {\bf  1984}, {\it 81(8)}, 3684-3690.
\bibitem{Parrinello-Rahman}
Parrinello, M.; Rahman, A. Crystal structure and pair potentials: A molecular-dynamics study. {\it Phys. Rev. Lett.} {\bf 1980}, {\it 45(14)}, 1196.
\bibitem{PME}
Essmann, U.; Perera, L.; Berkowitz, M. L.; Darden, T.; Lee, H.; Pedersen, L. G.  A smooth particle mesh Ewald method. {\it J. Chem. Phys.} {\bf 1995}, {\it 103(19)}, 8577-8593.
\bibitem{lincs}
Hess, B.; Bekker, H.; Berendsen, H. J.; Fraaije, J. G.  LINCS: a linear constraint solver for molecular simulations. {\it J. Comput. Chem.} {\bf 1997}, {\it 18(12)}, 1463-1472.
\bibitem{Gromos43a1}
van Gunsteren, W. F.; Billeter, S. R.; Eking, A. A.; H\"unenberger, P. H.; Kr\"uger, P.; Mark, A. E.; Scott, W. R. P.; Tironi, I. G. Biomolecular simulation: the GROMOS96 manual and user guide. 
Vdf Hochschulverlag AG an der ETH Z\"urich, {\bf 1996}, 86.
\bibitem{SPC3}
Wu, Y.; Tepper, H. L.; Voth, G. A. Flexible simple point-charge water model with improved liquid-state properties. {\it J. Chem. Phys.} {\bf 2006}, {\it 124(2)}, 024503.
\bibitem{KBUWFF}
Weerasinghe, S.; Smith, P. E. A Kirkwood-Buff Derived Force Field for Mixtures of Urea and Water. {\it J. Phys. Chem. B} {\bf 2003}, {\it 107(16)}, 3891-3898.
\bibitem{debashish2012}
Mukherji, D.; van der Vegt, N. F.; Kremer, K. Preferential solvation of triglycine in aqueous urea: An open boundary simulation approach. {\it J. Chem. Theory Comput.} {\bf 2012}, {\it 8(10)}, 3536-3541.
\bibitem{OPLS}
Jorgensen, W. L.; Maxwell, D. S.; Tirado-Rives, J. Development and testing of the OPLS all-atom force field on conformational energetics and properties of organic liquids. {\it J. Am. Chem. Soc.} {\bf 1996}, {\it 118(45)}, 11225-11236.

\bibitem{Kremer-Grest}
Kremer, K.; Grest, G. S. Dynamics of entangled linear polymer melts: A molecular-dynamics simulation. {\it J. Chem. Phys.} {\bf 1990}, {\it 92(8)}, 5057-5086.

\bibitem{espresso}
Halverson, J. D.; Brandes, T.; Lenz, O.; Arnold, A.; Bevc, S.; Starchenko, V.; Kremer, K.; Stuehn, T.; Reith, D. ESPResSo++: A modern multiscale simulation package for soft matter systems. {\it Comput. Phys. Commun.} {\bf 2013}, {\it 184(4)}, 1129-1149.
\bibitem{lammps}
Plimpton, S. Fast parallel algorithms for short-range molecular dynamics. 
{\it J Comp. Phys.} {\bf 1995}, {\it 117(1)}, 1-19.
\bibitem{kbi}
Kirkwood, J. G.; Buff, F. P. The statistical mechanical theory of solutions. I. {\it J. Chem. Phys.} {\bf 1951}, {\it 19(6)}, 774-777.
\bibitem{robin2016}
Cortes-Huerto, R.; Kremer, K.; Potestio, R. Communication: Kirkwood-Buff integrals in the thermodynamic limit from small-sized molecular dynamics simulations. {\it J. Chem. Phys.} {\bf 2016}, {\it 145}, 141103.
\bibitem{Petris}
Petris, P. C.; Anogiannakis, S. D.; Tzounis, P. N.; Theodorou, D. N.  Thermodynamic Analysis of n-Hexane-Ethanol Binary Mixtures Using the Kirkwood-Buff Theory. {\it J. Phys. Chem. B} {\bf 2018}, {\it 123(1)}, 247-257.

\bibitem{Urry}
Urry, D. W. Physical chemistry of biological free energy transduction as demonstrated by elastic protein-based polymers. {\it J. Phys. Chem. B} {\bf 1997}, {\it 101}, 11007-11028.

\bibitem{pnipmam}
Gernandt, J.; Frenning, G.; Richtering, W.; Hansson, P. A model describing the internal structure of core/shell hydrogels. {\it Soft Matter} {\bf 2011}, {\it 7(21)}, 10327-10338.
\bibitem{Ishihara}
Kiritoshi, Y.; Ishihara, K. Preparation of cross-linked biocompatible poly (2-methacryloyloxyethyl phosphorylcholine) gel and its strange swelling behavior in water/ethanol mixture. {\it J. Biomater. Sci. Polym. Ed.} {\bf 2002}, {\it 13(2)}, 213-224.
\bibitem{Ohkura}
Ohkura, M.; Kanaya, T.; Keisuke, K. Gels of poly (vinyl alcohol) from dimethyl sulphoxide/water solutions. {\it Polymer} {\bf 1992}, {\it 33(17)}, 3686-3690.
\bibitem{Hiroki}
Hiroki, A.; Maekawa, Y.; Yoshida, M.; Kubota, K.; Katakai, R. Volume phase transitions of poly (acryloyl-l-proline methyl ester) gels in response to water-alcohol composition. {\it Polymer} {\bf 2001}, {\it 42(5)}, 1863-1867.

\bibitem{pnipam2}
de Oliveira, T. E.; Netz, P. A.; Mukherji, D.; Kremer, K. Why does high pressure destroy co-non-solvency of PNIPAm in aqueous methanol? {\it Soft Matter} {\bf 2015}, {\it 11(44)}, 8599-8604.

\bibitem{mukherji15jcp}
	Mukherji, D; Marques, C. M.; Stuehn, T.; Kremer, K.
	Co-non-solvency: Mean-field polymer theory does not describe polymer collapse transition in a mixture of two competing good solvents.
        {\it J Chem. Phys.} {\bf 2015}, {\it 142}, 114903.

\bibitem{pnipam3}
Mukherji, D.; Wagner, M.; Watson, M. D.; Winzen, S.; de Oliveira, T. E.; Marques, C. M.; Kremer, K. Relating side chain organization of PNIPAm with its conformation in aqueous methanol. {\it Soft Matter} {\bf 2016}, {\it 12(38)}, 7995-8003.

\bibitem{ureadenature1}
Brandts, J. F.; Hunt, L. Thermodynamics of protein denaturation. III. Denaturation of ribonuclease in water and in aqueous urea and aqueous ethanol mixtures. {\it J. Am. Chem. Soc.} {\bf 1967}, {\it 89(19)}, 4826-4838.

\bibitem{mvalue}
Auton, M.; Bolen, D. W. Predicting the energetics of osmolyte-induced protein folding/unfolding. {\it Proc. Natl. Acad. Sci. U. S. A.} {\bf 2005}, {\it 102(42)}, 15065-15068.

\bibitem{rnetz}
Horinek, D.; Netz, R. R. Can simulations quantitatively predict peptide transfer free energies to urea solutions? Thermodynamic concepts and force field limitations. 
{\it J. Phys. Chem. A} {\bf 2011}, {\it 115(23)}, 6125-6136.

\bibitem{Berne}
Zhou, R.; Huang, X.; Margulis, C. J.; Berne, B. J. Hydrophobic collapse in multidomain protein folding. {\it Science} {\bf 2004}, {\it 305(5690)}, 1605-1609.

\end{thebibliography}
\end{document}